# An analysis of protesting activity and trauma through mathematical and statistical models

Nancy Rodríguez[1,*] and David White[2]

[1]University of Colorado at Boulder, USA
[2]Denison University, USA
[*]corresponding author, rodriguez@colorado.edu

December 31, 2023

## Abstract

The effect that different police protest management methods have on protesters' physical and mental trauma is still not well understood and is a matter of debate. In this paper, we take a two-pronged approach to gain insight into this issue. First, we perform statistical analysis on time series data of protests provided by ACLED and spanning the period of time from January 1, 2020, until March 13, 2021. We observe that the use of kinetic impact projectiles is associated with more protests in subsequent days and is also a better predictor of the number of deaths in subsequent deaths than the number of protests, concluding that the use of non-lethal weapons seems to have an inflammatory rather than suppressive effect on protests. Next, we provide a mathematical framework to model modern, but well-established psychological and sociological research on compliance theory and crowd dynamics. Our results show that understanding the heterogeneity of the crowd is key for protests that lead to a reduction of social tension and minimization of physical and mental trauma in protesters.

## 1 Introduction

The United States of America emerged as a free country on August 2, 1776, as a result of protests against England's colonial rule. Interestingly, in spite of this history, its government and police departments have had a fraught relationship with protesters [12]. While the right to protest is protected by the First Amendment, police authorities have historically played a significant role in managing protests, even when they are peaceful demonstrations. Nonetheless, protests continue to be a powerful example of effective collective behavior that can lead to a clear change in the establishment. A notable example is that of the civil rights movement of the 1960s [38], and a more recent example is the wave of protests that overtook the country in support for black lives in 2012 after the death of Trayvon Martin [26]. Throughout the years, police departments across the United States have used a variety of protest control strategies, which will be discussed in detail in section 2.2. A natural question that has inspired much research is that of which strategy is the most effective for managing protests, see [12, 45, 28] and the references therein. While this question is far from being answered, evidence from an assessment ordered by the Justice Department's Office of Community Oriented Policing Services concluded that, in the recent protests/riots that were triggered by the killing of Michael Brown in Furgeson, Missouri by a white police officer in 2014, the police strategy of showing dominant force in an effort to get protesters to comply only escalated the violence [17].

The reasons that bring individuals from all walks of life to gather and protest in mass are varied. However, protests can be interpreted as a sign of high *social tension* in a system. According to Parson *social tension* is the tendency to disequilibrium in the balance of exchange between two or more components in a system [35], and accumulated social tension can reveal itself in protest action [19]. Thus, on the one hand, properly managed protesting activity can lead to reduced social tension, when properly controlled, by clearing the air [9]. On the other hand, poorly controlled protests can lead to increased tension, escalated violence, and a further disequilibrium of the system. In fact, there have been many instances of significant physical and mental trauma resulting from violent events that occur during protests and riots [20, 31, 21]. This results in a collective trauma that affects not only the individuals experiencing the violence directly but also those who witness it [31, 18]. While the physical injuries occurring in such events are observable and trackable in theory, in practice there is no universal method used to keep track of these data. Instead, we rely on doctors and hospitals to monitor and share details of relevant injuries [43]. Mental trauma is even harder to measure as individuals may not know they are experiencing post-traumatic stress disorder, or a similar mental health condition, and may not seek treatment. Nevertheless, these pose a significant health problem to our society and should not be ignored.

The interaction between protesters and police is complex, particularly because of the competing effects at play. On the one hand, police presence may deter some individuals from committing violent or acts of aggression (AofA), especially when counter-protesters may be present. On the other hand, police can escalate violence or even turn a peaceful demonstration violent [17]. Generally, it is difficult to determine which of the competing effects is dominating. The question motivating us is thus: Does police presence mitigate or escalate violence at protests? Our two-pronged approach uses both statistical and dynamic models. We first perform a time series analysis on data from The Armed Conflict Location & Event Data Project (ACLED) database [37], spanning the period of time from January 1, 2020, until March 13, 2021. This publicly available database is maintained on the ACLED website by experts and is based on a variety of media sources. This data set has events, classified as protests, and riots, to name a few, and contains information such as the date, location, actors involved (e.g., protesters, military, police, etc.), the number of fatalities, etc. These data allow us to explore the effect that things like current events, police involvement, and fatalities have on future events. However, these data lack the depth to be able to fully understand the effect that different police management methods have on the collective trauma of protesters. To accurately tease out the effect of protest policing strategies, accurate, detailed, and transparent data needs to be collected, which is vital for data-driven modeling. In the absence of data, mathematical modeling of relevant sociological theories can offer an avenue for us to understand the effect that different protest management strategies have on acts of violence and aggression. In [28], Maguire argues for a new direction in protest policing based on current psychology/sociology research on (1) compliance theory and (2) crowd dynamics. Here we present and analyze a dynamical systems model based on the psychology research discussed in [28]. We note that it is difficult to parameterize these models given the current data available. The value of the model introduced here is that it helps us quantify the effects of different policing strategies under the point of view championed in [28]. These are *thought experiments* that can help us gain insight into how crowd heterogeneity and protest management strategies have an effect on the collective trauma due to protests. Our work builds on a preliminary analysis [24], that focused on the effect of police use of kinetic impact projectiles on protest dynamics.

## 1.1 Previous work

Protests are complex social events that have been pervasive throughout history and which have been the subject of intensive research activity, see for example [7, 25, 10]. A statistical lens has been taken in [15] where the authors categorize five kinds of police responses to protests, ranging from "police do not show up" to "use of arrests and barricades" to "use of tear gas, etc." and fits a time-series model to the data (albeit with a few statistical flaws). The paper finds that the impacts of police repression techniques tend to be short-term, e.g., one week. In [11] the authors analyze the effect that the US Department of Defense 1033 program, which allows the Department of Defense to give state, local, and federal law enforcement agencies military hardware, has on violent behavior among officers. They find a positive and statistically significant relationship between the number of 1033 transfers and fatalities from officer-involved shootings across all of their models.

There is also a small body of literature studying the mental health effects of exposure to macro-level trauma, such as protests. In [16] the authors look at post-traumatic stress (PTS) symptoms caused by the fatal shooting of Michael Brown in August 2014 and subsequent civil unrest in Ferguson, MO. The study found a direct association between the factors of race and protest engagement with PTS symptoms. Media exposure to these events was indirectly associated with PTS symptoms. The authors of [? ] did a meta-analysis of the effect of collective action on mental health. The study collected papers from PubMed, Web of Science, PsycINFO, and CINAHL Plus from their inception until January 1, 2018. One of the factors found to be linked to poorer mental health after participation in a protest was exposure to violence. The authors also point out two studies that showed that collective action could lead to lower levels of depression and suicide, possibly due to the feeling of belonging and social cohesion.

Collective protesting behavior has also been modeled with mathematical models. In [10] Davies and collaborators developed a mathematical model for the 2011 London riots and their policing, taking into account the contagious nature of participation; the distances traveled to riot locations; and the deterrent effect of policing. In [4, 5], the authors introduced a model studying the evolution of protesting activity coupled with social tension.

The mathematical approach of this work bears similarities to recent literature on the use of mathematics in the analysis of uncivil and criminal activities [40, 4, 39, 5], showing for instance that patterns, which are useful to understand, emerge when looking at the macroscopic scale.

## 2 Sociological background

### 2.1 Social tension and conflict

Social tension and conflict often go hand-in-hand: social tension precedes conflict but does not always lead to conflict [1]. In the sociological literature, social tension is generally defined as pressure felt by a collective. This pressure can result from things like rupture of social ties, increased social anomie [34], or a buildup of mental fatigue and irritability, frustration and deprivation, aggression, and depression of a significant part of society [42]. Social tension, an integral part of all social systems, can spread among the population and manifests itself in mass actions [8]. According to Cosner, conflict can serve the very important role of regulating systems that are out of equilibrium, and socially-controlled conflict can remove accumulations of suppressed hostile emotions, thus reducing social tension [9].

Social tension is recognized to be a natural part of society and while it generally has a negative connotation, it can also play a positive role. Mainly, this quantity can be used to track the potential for conflict [1]. While conflict, when managed properly, can be productive, it should be avoided when possible. Thus, tracking social tension at its different stages and making changes in the system to reduce social tension, before conflict arises, would be ideal. Researchers have described different developmental stages of social tension. The initial stage is latent when a negative emotional state begins to emerge in a population without their perception. The second stage is the perception and spread of these negative emotions. These negative emotions can reach a level in a population that manifests itself in conflict or protesting activity [1].

### 2.2 Policing strategies background and history

To fully understand the dynamics of protests in the United States one must understand the interaction between police and protesters. The individual actions of each actor, either police or protesters, modify the environment of the other and thus potentially alter the others' actions ([12], Part I, Chapter II). In the United States, police departments have a history of managing protests. This is evidenced by the development of the permitting system used initially by the three major Washington, D.C., police agencies (the National Park Service Police, the U.S. Capitol Police, and the Metropolitan Police of the District of Columbia), which have been replicated by other state capitals ([12], Part I, Chapter II). It is evident that outcomes of protests depend not only on the protesters' intentions but also on their interaction with the police (when present). Therefore, we must understand the history of policing strategies and their potential effect on protest dynamics. In Chapter 2 of [12] the authors introduce five key characteristics in protest policing practices. These are:

1. The extent of police *concern regarding the First Amendment rights of protesters*, and police obligations to respect and protect those rights;
2. The extent of police *tolerance for community disruption*;
3. The nature of *communication* between police and demonstrators;
4. The *extent and manner of arrests* as a method of managing demonstrators;
5. The *extent and manner of using force* in lieu of or in conjunction with arrests in order to control demonstrators.

It is natural to begin in the 1960s, where the dominant model was an **escalated force** model. In the escalated force model, police disregard or minimize the First Amendments rights of protesters, do not tolerate any forms of disruptions, have limited communication with protesters, and are quick to arrest protesters (even if laws have not been broken), and use force as a standard method to deal with protesters. Historically, the use of this method led to many arrests, beatings, and deaths of protesters [12], and thus activists called for the introduction of "less-lethal" tools, such as Kinetic Impact Projectiles (KIPs). In that era, such tools probably saved lives, since a rubber bullet is less lethal than a bullet [12].

In the 1980s and 1990s, the **negotiated management** model became the dominant model. In complete contrast to the escalated force model, two primary goals of the negotiated management model were to protect a protester's First Amendment rights and to save lives. Under this management style, some disruptions were anticipated and a high level of communication between police and protesters was expected. Moreover, arrests and the use of force were only used as a last resort [28]. In this management model, Kinetic Impact Projectiles were unnecessary. In the late 1990s, the negotiated management model was replaced by the more aggressive approaches known as the *strategic incapacitation model*, *command and control model*, and the *Miami model* [28]. These are all variations of the escalated force model.

## 2.3  Deterrence theory versus procedural justice

The "show of force" policing strategy used during the protests/riots that ensued after the 2014 killing of Michael Brown, is based on *deterrence theory*, stating that individuals either comply or become defiant based on the cost associated with these actions [36]. The premise of this theory is that when the police show a forceful presence, for example, starting from the attire that they wear, the arrests that they make, or a more violent show of force, the cost of being defiant will be too high [36, 28]. An alternative and more modern theory is that individuals make their decisions to comply or become defiant based on what they perceive to be *fair* [22, 46, 30]. This theory goes by the name of *procedural justice*, and is concerned with how people perceive the fairness of the procedures used by an authority figure. Figure 1b illustrates the four components that are necessary for an interaction between a law enforcement agent and an individual to be deemed fair. There is research evidence that people's perceptions of procedural justice influence their judgments about the legitimacy of law enforcement [44]. Under the assumption of a process-based model of regulation when law enforcement agents are perceived to behave in an unjust way, "the police and the law are viewed as less legitimate, and people are less likely to comply" [28] Under the assumption of this theory, any actions by the police that are perceived by the crowds to be unfair will escalate the tension.

A body of research from the 2010s suggests that the "negotiated management" model and de-escalation tactics are most effective, both at controlling protests and preventing violence between protesters and police. For example, an after-action assessment sponsored by the Justice Department's Office of Community Oriented Policing Services after the killing of Michael Brown found that the heavy-handed police response to protests relied on "ineffective and inappropriate strategies and tactics" that had the "unintended consequence of escalating rather than diminishing tensions." [28, 33]

## 2.4 Social identity theory

Protest policing strategies are heavily developed based on crowd psychology. The classical view of how individuals act when part of a crowd dates back to the French scholar Gustave Le Bon's 1895 book *The Crowd: A Study of the Popular Mind*. From Le Bon's point of view, when an individual joins a crowd, they lose their identity and become an "automaton who ceased to be guided by his will." A different and more modern perspective comes from *social identity theory*, which refers to "the way in which people understand how they are positioned relative to others" [23]. From this perspective, it is important for authorities to understand the various social identities in a protest in other to use effective and fair policing strategies [28]. Social identity theory later evolved into the *elaborated social identity model* (ESIM), which allows for the possibility of individuals to shift their social identity temporarily [13]. It has been argued that treating a group of protesters as homogeneous is extremely dangerous [28]. Nowadays, many protests consist of people from all over the country with a variety of social identities. The ESIM allows for individuals with different social identities to gather and take on a different identity while protesting [28].

An ESIM perspective posits that ill-advised actions by police can instigate or escalate conflict and violence in crowds [28, 13]. In evaluating police response as unjust, a protest participant will be more likely to defy police authority or even use violence. It may be the case then that protest participants "will unite around a sense of opposition to the police and the authorities they are protecting [28, 13]."

## 2.5 Productive action

From the sociological theory discussed above, we observe that protesting activity can lead to a reduction of social tension when the management of the protest is adequate. Figure 1a illustrates a Venn diagram summarizing the discussion from above. In particular, we see that protesting action is a consequence of high social tension and a triggering event. If protest participants feel that they have been treated in a procedurally just manner, then protesting can lead to productive action, which we define to be a reduction of tension and the minimization of physical and mental trauma experienced by the protesting group. Our aim then is to understand what protest management models lead to such productive action.

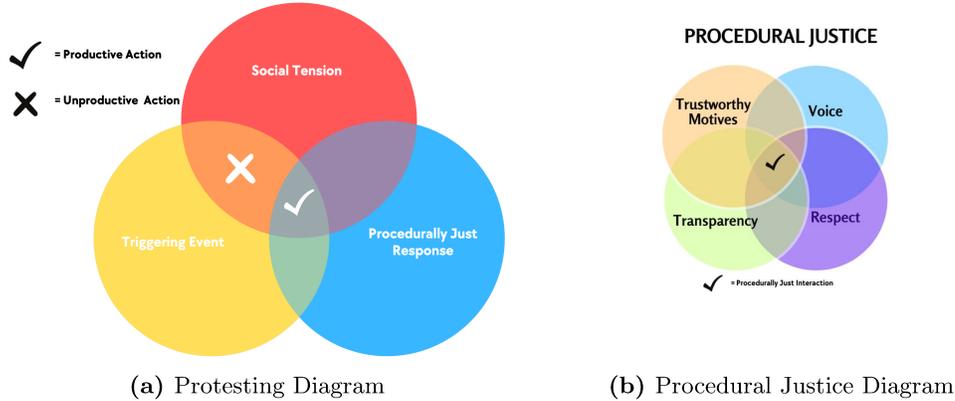

**(a)** Protesting Diagram  **(b)** Procedural Justice Diagram

**Figure 1:** Diagrams illustrating the sociological theories that we aim to provide a mathematical framework for: (a) the factors necessary for productive action and (b) the four components needed for an interaction to be deemed procedurally just.

## 3 Methods

### 3.1 Data acquisition and cleaning

We analyzed data from The Armed Conflict Location & Event Data Project (ACLED) database, spanning the period of time from January 1, 2020, until March 13, 2021. This publicly available database is maintained on the ACLED website [37], is based on a variety of media sources, and is maintained by experts. Each row is an event (e.g., a protest, riot, or strategic development), and 29 columns contain information related to the type of event, date, location, actors involved (e.g., protesters, military, police, etc.), the number of fatalities, and a 'notes' field describing the event.

We curated the data to focus on protests in the USA. We then employed text processing algorithms to identify (based on the 'notes' field) which protests involved 'kinetic impact projectiles' (KIPs), i.e., police use of rubber bullets, foam rounds, bean bag rounds, or tear gas. We then wrangled the data into a new data frame where each row is a day, with columns of time series telling how many protests happened that day, how many fatalities were associated with the protests that day, and how many protests involved KIPs. A similar analysis was carried out in [2, 27].

### 3.2 Statistical analysis

We carried out a cross-correlation analysis to determine which of these time series led/lagged the others, and we fit ARIMA models to determine how each of these time series depends on its own past. Lastly, we fit multivariate models, e.g., predicting the number of protests based on KIP usage. This allowed us to quantify the impact of a single protest involving KIPs, on the overall number of protests in subsequent days. We used an information-theoretic measurement, the Akaite Information Criterion (AIC) to decide between models. The model with the lowest AIC is most likely to correctly forecast future data [41].

We now give a brief summary of ARIMA models, following [41]. An ARIMA model consists of three parts: an autoregressive (AR) part, an integrating part, and a moving average (MA) part. Before the AR and MA parts can be fit, the time series must be stationary, i.e., the mean, variance, and autocorrelation structure should not change over time. If the given time series $x_t$ is not stationary then we apply the differencing operator $\Delta x_t = x_t - x_{t-1}$ to make it stationary. The $\Delta$ operator can be iterated if necessary. The "integrating part" of an ARIMA model tells how many times the differencing transformation was applied. Once our transformed variable (we will denote it $X_t$) is stationary, we use autocorrelation plots, partial autocorrelation plots, and lagplots (which show us the correlation between $X_t$ and its lags $X_{t-1}, X_{t-2}, \ldots$) to determine the AR and MA parts, which determine how many lags of $X_t$ and of the error term $\epsilon_t$ we should include in the model. For example, an ARIMA(3,1,2) model means we work with $X_t = \Delta x_t$, and we model $X_t$ as a linear function of three AR lags $X_{t-1}, X_{t-2}, X_{t-3}$ and two MA error lags $\epsilon_{t-1}$ and $\epsilon_{t-2}$. That is, we write

$$X_t = \beta_0 + \beta_1 X_{t-1} + \beta_2 X_{t-2} + \beta_3 X_{t-3} + \alpha_1 \epsilon_{t-1} + \alpha_2 \epsilon_{t-2} + \epsilon_t.$$

If $\beta_1 > 0$ and if $X_t$ was the number of protests on the day $t$, then this model tells us that, after a large day of protests, we can expect an even bigger number of protests the next day. If $\alpha_1 > 0$ it means that, if yesterday had an unusually large number of protests, then we can expect today to have a larger number. Certain ARIMA models represent self-exciting time series.

A seasonal ARIMA (or SARIMA) model further allows seasonal lags like $X_{t-7}$, the number of protests seven days ago, or $\epsilon_{t-7}$, a shock event from last week. The goal, in the end, is for the vector of residuals, $\epsilon_t$, to be random and independent, meaning it exhibits no further autocorrelation. In particular, the correlation between $\epsilon_t$ and each $\epsilon_{t-h}$ should be statistically insignificant according to the autocorrelation function and Ljung-Box test.

Applying this modeling framework, we found that the best model for the number of protests was a SARIMA(3,1,2)x(2,0,0)[7], i.e., after first-order differencing, $X_t$ depends on each of the past three days, and whether the past two days were unusual, and also depends on seven days ago and fourteen days ago (due to the seasonal AR(2) part).

Differencing was required because the time series failed to have a constant mean and variance over time. In the summer of 2020, the number of protests spiked.

Now let $p_t$ (respectively $k_t$, resp. $d_t$) denote the number of protests (resp. protests with KIP usage, resp. fatalities) on the day $t$. Our cross-correlation analysis showed strong positive correlations between $p_t$ and $k_t, k_{t-1}, k_{t-2}, k_{t-3}, k_{t-4}$, and $k_{t-5}$. This suggests that KIP usage is associated with increased numbers of protests in the subsequent days. These correlations were statistically significant at lags 0, 1, and 3.

Similarly, $d_t$ was positively associated with $k_t$, $k_{t-1}$, and $k_{t-2}$, meaning that KIP usage is associated with more deaths in the subsequent days. These correlations were statistically significant at lags 0 and 1.

Next, we fit models to predict $p_t$ based on its own past and based on $k_t$. The best-fitting ARIMA model (chosen by AIC) for protests as a function of KIP use involves the number of protests on the previous day, the number of protests seven days earlier, the number of KIPs used, and the error terms from the previous day and the day before that. Controlling for the effect of the past, every protest in which KIPs were used was associated with 22.5 more protests, and this is statistically significant ($p < 0.001$). We note that, due to a lack of normality of the residuals, this $p$-value was obtained nonparametrically, using randomization-based inference. When including each of the days in the past two weeks, each KIP usage was associated with 19.32 more protests the next day ($p < 0.001$), and 15.9 more protests a full week after the KIP usage.

Lastly, we fit models to predict $d_t$ based on its own past and based on $k_t$. We find that each use of KIPs was associated with 0.134 more deaths the same day, and 0.129 more deaths the next day (both with $p < 0.001$). Furthermore, KIP use was a better predictor than the number of protests, for deaths.

## 3.3 A dynamics perspective

To better understand the complex interactions between the police managing a protest and the protesters themselves, we develop a mathematical model based on the sociological theory discussed above. In particular, we assume that protesters evaluate interactions with police from a procedural justice lens. Moreover, moving away from Le Bon's point of view, we assume that protesters have heterogeneous backgrounds and intentions and that they can shift their intentions dynamically based on police-protester interactions. The mathematical framework we develop is based on evolutionary game theory, which has been widely used to study animal and human behavior, see [3] and references therein. In this framework, one can observe players employing different strategies based on the different interactions that occur. Our goal is to provide a way to quantify the effect of different protest management strategies under the assumptions discussed in § 2. In particular, this framework allows us to see the dynamics of protests unfolding as time evolves.

We assume that a protest has two types of actors (or players): police agents and protesters. Within the protester groups, we consider that there are subgroups with different values [14]. For simplicity, we subdivide our protesters into *agitators* and *moderates*. In this way, we can tease out the effect of subgroups with different values and intentions on how the protests unfold in a simple setting. Such a split is common in the conflict sociological literature, see for example [32]. We assume that agitators have some probability of perpetrating an act of aggression, which could be something as simple as throwing a water bottle or something more violent. These agitators could be individuals who truly support the cause and are open to becoming aggressive if the situation calls for it. However, this group could also have members who do not support the cause and are simply there to create havoc. For example, in some Black Lives Matters protests, members of white supremacist groups show up to agitate the crowd [6]. Another interesting example where this occurred was during the Revolution of Dignity in Ukraine from late 2013 to early 2014 when the Ukrainian government paid militia to attend protests and incite violence [29].

Police and protesters make different choices as the progress of the protest. Police make the choice of whether or not to commit an act of aggression. Their decisions are informed by the management model in place (we consider the escalated force model versus the negotiated management model) and also by the choices made by the protesters. Agitators also make the choice of whether or not to commit an act of aggression. This choice is informed by the number of acts of aggression (AofA) committed by the police and also by fellow protesters. In essence, we are assuming that an escalation of violence or acts of aggression occurs. Moderates do not perpetrate any acts of aggression but do have a choice of whether or not they will become an agitator. In making this choice, moderates observe the choices that police make and judge them using a procedural justice lens. All protesters have the choice to leave or stay at each time step.

## 3.4 An evolutionary game theoretic model

Let us assume that there are $N$ protesters at the beginning of a given protest. Let $N_a(t)$ and $N_m(t)$ denote the number of agitators and moderates, respectively. For simplicity, we assume that $N_a(t) + N_m(t) \leq N$ for all time, so that no new protesters arrive, but protesters eventually leave. In reality, there are certainly cases where new individuals join a protest. This could be included in the model with ease; however, we seek the most parsimonious model that will allow us to tease out the effects of different protest management models, and including an influx of new protesters will not change the dynamics (except to potentially increase magnitudes of the acts of aggression or prolong the protest). Denote the number of agitators at time $t$ by $u_1(t)$, the number of moderates by $u_2(t)$, and the number of police present at time $t$ by $p(t)$. As time evolves we keep track of the number of acts of aggression (AofA) perpetrated by the agitators that have occurred by time $t$ and denote this quantity by $v_1(t)$. Moreover, we keep track of the number of acts of aggression or arrests made by the police that have occurred by time $t$ and denote the quantity by $v_2(t)$. Finally, we denote the social tension in the system at time $t$ by $\tau(t)$, or just $\tau$ if $t$ is clear from the context. We summarize these variables and their description in Table 1.

### 3.4.1 The choices of protesters

In light of the above discussion, we assume that in making the choice of whether or not to commit an act of aggression, agitators will consider competing factors. On the one hand, police presence will provide a deterrence effect, while high social tension will increase the likelihood of committing an act of aggression. Thus, we assume that the probability that an aggressive act is committed by a particular agitator during a period of time $\delta t$ is given by:

$$P_{u_1}(\delta t) = 1 - e^{-\left(\frac{f_1(\tau)}{p(t)+1}\delta t\right)} \quad \text{and} \quad f_1(\tau) = \begin{cases} 0, & \tau < \tau_c, \\ T_1, & \tau \geq \tau_c, \end{cases} \quad (1)$$

where $\tau_c$ represents a critical threshold of the social tension required for an aggressive act to occur and $T_1$ the intensity. The function $f_1$ is a step-function modeling an all-or-nothing response. In reality, individuals have different thresholds of social tension necessary to commit an act of aggression. We see $\tau_c$ as a "mean-field" parameter governing the choices of the populations of agitators as a whole. We will analyze what happens when $\tau_c = 0$, which corresponds to the case where agitators will commit acts of aggression independent of the social tension. Another suitable choice for $f_1$ could be a Sigmund-type function. We also point out that we view arrests as acts of aggression even if they are "justified". This is a simplification given that some protesters may view certain arrests as warranted. However, this is mitigated by how moderates perceive fairness and make decisions to either stay a moderate, become an agitator, or leave the protest. In particular, moderates have the choice of becoming agitators and they make this decision by evaluating the fairness of protester-police interactions. Specifically, if they perceive the police management to be unfair, they will be more likely to become agitators. One way to measure fairness is to consider the fraction of acts of aggression that are perpetrated by the police relative to the total number of acts of aggression. In other words, if moderates see that the protesters are being too destructive and violent, they may deem arrests and police acts of aggression as warranted. On the other hand, if the police are too heavy-handed, the moderates may be more likely to join the agitators. We choose to measure this by the quantity $\frac{v_2}{v_2+v_1+1}$, where the one in the denominator is to avoid division by zero. Hence, the probability that a moderate turns into an agitator during the time interval $\delta t$ is:

$$P_{u_2 \to u_1}(\delta t) = 1 - e^{-\frac{v_2}{v_2+v_1+1} f_3(\tau)\delta t}, \quad (2)$$

where the function $f_3$ has a similar form to $f_1$, but with potentially a different critical threshold. At any given time in the protest, all protesters have a choice to exit the protest and they do so at a rate denoted by $\epsilon$.

### 3.4.2 The choice of the police agents

The choice that police agents have is whether or not to commit an act of aggression directed toward the protesters or to arrest protesters. As mentioned earlier, we aggregate these two and refer to both actions as acts of aggression. We assume that the probability of an act of aggression being committed by the police depends on the number of acts of aggression taken by the protesters, which is the quantity $v_1$. Thus, we arrive at the following term:

$$P_{u_2}(\delta t) = 1 - e^{-f_2(v_1)\delta t} \quad \text{and} \quad f_2(v) = \begin{cases} 0, & v < v_c, \\ T_2, & v \geq v_c, \end{cases} \quad (3)$$

where $v_c$ represents a critical number of acts of aggression that must occur before the police engage in the use of force and/or make arrests, and $T_2$ is the intensity. In this work, we consider the *escalated force* and the *negotiated management* models as the two main protest policing strategies. The differences in the models lead to different model parameters.

The **escalated force model** is characterized by little to no tolerance of disorder of the protesters, and police are quick to arrest. Thus, the threshold $v_c$ is zero or something very small. Moreover, there is no communication with protesters and little to no respect for protesters' First Amendment rights. We assume that police presence thus leads to an increase in social tension. Even some police acknowledge that deploying police in riot gear can backfire. Some quotes from police officers found in [28] that are of interest are: in relation to officers wearing riot gear, "it incites a type of reaction that might backfire and agitate;" "once the hats and bats and turtle suits come out, it brings aggression with it."; "if you line up a bunch of police officers with riot gear and shields, you are telling protesters 'to go ahead and throw rocks and bottles at us'."

On the other hand, the **negotiated management model** is characterized by a much higher tolerance towards disorder and acts of aggression by the protesters. In that model, arrests and the use of force is a last resort. Moreover, there is a great deal of communication and respect towards protesters. Thus, in this model, the threshold $v_c$ is much higher and police presence can help diffuse tensions since the communications between police and protesters are open.

### 3.4.3 The evolution of the social tension

As described in the introduction, social tension precedes protesting activity and conflict. The goal is to have productive action that can lead to a decrease in social tension with a minimum amount of physical and mental trauma, both for police and protesters. We make the assumption that, regardless of the police management strategy being used, the social tension increases with the total number of acts of aggression, $v_1 + v_2$, and has a natural decay with time:

$$\tau(t + dt) = \tau(t) + \theta(v_2(t + \delta t) + v_1(t + \delta t) - v_2(t) + v_1(t)) - \omega \tau(t), \tag{4}$$

where $\omega$ is the rate of decrease if the number of aggressive acts remains static and $\theta$ the increase in social tension per act of aggression. The difference in communication strategies for the two management models that we consider can be modeled in the parameter $\theta$. In particular, a situation where police are in communication with protesters, e.g., asking protesters about the reasons they are participating, can help attenuate the effects of acts of aggression, leading to a lower $\theta$. On the contrary, a lack of communication can enhance the effect that an act of aggression has on social tension, leading to a higher $\theta$.

| Notation | Description |
|---|---|
| $u_1(t)$ | Total number of agitators at time $t$. |
| $u_2(t)$ | Total number of moderates at time $t$. |
| $p(t)$ | Total number of police at time $t$. |
| $v_1(t)$ | Total number of protester AofA at time $t$. |
| $v_2(t)$ | Total number of police AofA at time $t$. |
| $\tau(t)$ | Social tension level at time $t$, sometimes denoted $\tau$. |
| $\tau_c$ | Critical tension needed for agitators to become violent. |
| $v_c$ | Number of AofA perpetrated by protesters for police to engage. |
| $T_1$ | Probability modulators of the number of AofA of protesters. |
| $T_1$ | Probability modulators of the number of AofA of police. |
| $\epsilon$ | Exit rate of protesters. |
| $\omega$ | Rate of decrease of tension. |
| $\theta$ | Increase of tension per act of violence. |

**Table 1:** Description of the notation in the evolutionary game theory.

**The time evolution of protesters and acts of aggression**

The flowchart in Figure 2 illustrates how a given protest evolves. Specifically, it shows the choice-making process of the protesters depending on which category they belong to, and how the system is updated depending on these choices. We can thus develop updated rules for this protest participation game as time evolves. Using the probability $P_{u_1}(\delta t)$ that a given agitator commits an act of aggression during a time period of size $\delta t$, we obtain that the total number of acts of aggression perpetrated by agitators by time $t + \delta t$ is given by:

$$v_1(t + \delta t) = v_1(t) + u_1(t)\left(1 - e^{-\left(\frac{f_1(\tau)\delta t}{p+1}\right)}\right). \tag{5}$$

From equation (5) we observe that as the tension increases, the number of acts of aggression perpetrated by the agitators also increases. On the other hand, there is a deterrence effect at play as the presence of more police members lowers the probability of an act of aggression happening. The competition between tension, which enhances acts of aggression, and police presence which inhibits them, is modulated by the parameter $T_1$. Specifically, $T_1 << 1$ implies that deterrence is the dominant factor, $T_1 >> 1$ implies that the tension dominates the dynamics, and $T_1 \approx 1$ balances the two effects. In a similar fashion, we can obtain the updating rule for the total number of acts of aggression perpetrated by police members by time $t+\delta t$. Using the probability that a particular law-enforcement agent commits an act of aggression, given by equation (3), we obtain the equation:

$$v_2(t+\delta t) = v_2(t) + p(t)\left(1 - e^{-f_2(v_1)\delta t}\right). \tag{6}$$

As we see from equation (6) the higher the variable $v_1(t)$, the more likely that a police agent will commit an act of aggression. Also, recall that there is a critical threshold, $v_c$, separating regions of zero acts of aggression perpetrated by the police agents and an escalating amount of acts of aggression perpetrated by the police. We equate an escalated force management model with small values of $v_c$ and a negotiated management model with large values of $v_c$.

The number of agitators at time $t + \delta t$ depends on how many exit the protest and how many moderates become agitators. For simplicity, we assume that all agitators who commit an act of aggression either decide to leave the protest or are arrested. This modeling choice has precedent; in [40] criminal agents are assumed to lay low for a while after committing a crime. Another way to interpret this modeling assumption is that we count all of an individual's acts of aggression as one. With this in mind, we obtain the updating rule given by:

$$u_1(t+\delta t) = u_1(t) + u_2(t)\left(1 - e^{-\frac{v_2}{v_2+v_1+1}f_3(\tau)\delta t}\right) - u_1(t)\left(1 - e^{-\left(\frac{f_1(\tau)}{p+1}\delta t\right)} + \epsilon\delta t\right). \tag{7}$$

Here and below, we write $v_1$ instead of $v_1(t)$, etc., to streamline the formulas. Since moderates either become agitators during a period of size $\delta t$, with probability given by (2), or exit at rate $\epsilon$, we obtain the updating rule for moderates:

$$u_2(t+\delta t) = u_2(t) - u_2(t)\left[\left(1 - e^{-\frac{v_2}{v_2+v_1+1}f_3(\tau)\delta t}\right) + \epsilon\delta t\right]. \tag{8}$$

## 3.5 Dynamical System Approximation

In this section, we approximate the evolutionary game theory described above by its dynamic systems counterpart. By taking the limit as $\delta t \to 0$ and linearizing all probabilities in equations (4)-(8) and we obtain a system of five coupled and nonlinear ordinary differential equations for the probability density functions of each of the variables, which is as follows:

$$\begin{cases} \frac{dv_1}{dt} &= u_1(t)\left(\frac{f_1(\tau)}{p+1}\right), \\ \frac{dv_2}{dt} &= p(t)f_2(v_1), \\ \frac{du_1}{dt} &= u_2(t)\left(\frac{v_2}{v_2+v_1+1}f_3(\tau)\right) - u_1(t)\left(\frac{f_1(\tau)}{p(t)+1} + \epsilon\right), \\ \frac{du_2}{dt} &= -u_2(t)\left(\frac{v_2}{v_2+v_1+1}f_3(\tau) + \epsilon\right), \\ \frac{d\tau}{dt} &= \theta\left[u_1(t)\left(\frac{f_1(\tau)}{p+1}\right) + p(t)f_2(v_1)\right] - \omega\tau(t). \end{cases} \tag{9}$$

A natural assumption on $p$ is that it is eventually zero. For example, if there are no protesters then it should hold that $p = 0$. We thus assume that:

(A1) $0 \leq p$ for all $t$ and $p(t) = 0$ if $u_1(t) + u_2(t) = 0$.

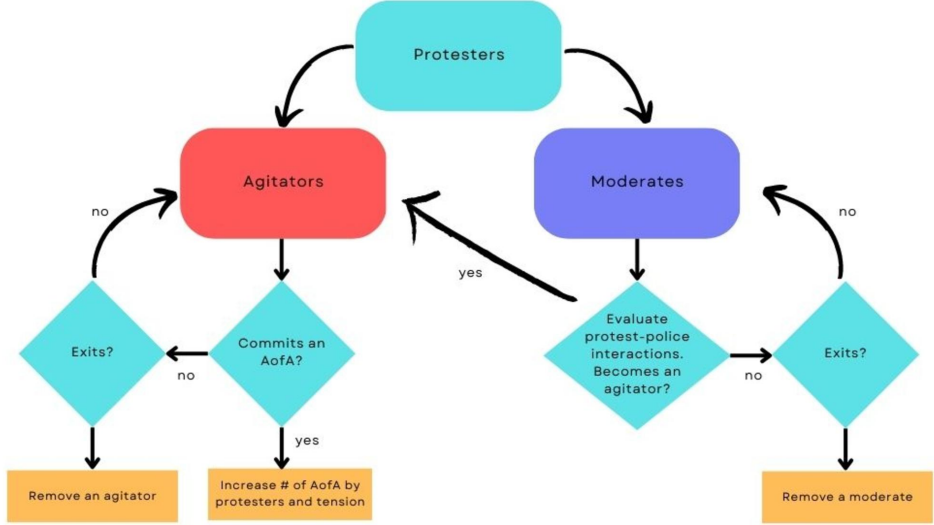

**Figure 2:** Protesters activity flow-chart

### 3.5.1 Estimates of the solutions

Note that the number of protesters will eventually be less than one, due to the flux being negative. Of course, because the system models the dynamics of a continuum population, the number of protesters will approach zero, but never achieve it. Moreover, condition (A1) guarantees that the number of police will eventually be zero. In fact, the third and fourth components of the solution satisfy the upper bounds:

$$u_2(t) \leq u_2(0)e^{-\epsilon t} \text{ and } u_1(t) \leq e^{-\epsilon t}\left(u_2(0)t + u_1(0)\right).$$

This implies that $\frac{dv_1}{dt} \to 0$ as $t$ increases and the number of acts of aggression committed by the protesters will taper off. However, if the police begin committing acts of aggression, the tension will increase and could eventually lead to agitators becoming aggressive. One can also see that under the following conditions, there will be zero acts of aggression:

(i) $\tau(0) < \tau_c$, $v_1(0) = v_2(0) = 0$.
(ii) $u_1(0) = 0$ and $v_1(0) = v_2(0) = 0$.

On the other hand, the following conditions guarantee positive $v_1$ and $v_2$ for $t > 0$.

(i) $\tau(0) > \tau_c$, $u_1(0) > 0$.
(ii) $\tau(0) < \tau_c$ and $v_2(0) > 0$.

## 4 Results

In this section, we analyze system (9). First, we study and illustrate the wide range of behaviors of the solution depending on model parameters. We are particularly interested in case studies analyzing the effect of different protest management models in homogeneously moderate crowds and the effect of heterogeneous crowds.

## 4.1 Solution dynamics and sensitivity analysis

In this subsection, we study the dynamics of the solutions to equation (9). Note that the exit rate of protesters and the natural decay of the tension have an effect on the duration of the protest. For the following analysis we choose to set $\epsilon = .02$ and $\omega = .01$. Moreover, we fix $f_3(z) = \frac{1}{10}\mathbb{1}_{z>2}$, where $\mathbb{1}$ is the characteristic function.

It is of interest to consider case studies with different initial conditions and to study the sensitivity of the system to the remaining parameters. Specifically, we perform local and global sensitivity analysis for parameters $T_1, T_2, \theta, \tau_c$, and $v_c$. To accomplish the global sensitivity analysis we use the R function senseRange(), where a distribution is defined for each sensitivity parameter and the model is run a large number of times, each time drawing values for the sensitivity parameters from their distribution. Here we assume a uniform distribution for all parameters within certain physically relevant ranges that are shown in Table 2. The senseRange() function produces envelopes around the sensitivity variables, see Figure 3 for an example. The turquoise envelope represents the range from the minimum values of the solutions up to the maximum values of the solutions. The blue envelope is the average dynamics plus/minus a standard deviation (which is why the dynamics dive into a negative regime). The local sensitivity analysis is done using the R function sensFun(), which estimates the local effect of the parameters on unknowns. This is done by calculating a matrix of sensitivity functions, which are the rates of change of the unknowns with respect to the parameters.

### 4.1.1 Case study 1: Homogeneously moderate crowd

We first consider the situation where there is a homogeneously moderate protest. In other words, there are no agitators in the crowd initially. Particularly, we consider two sets of initial conditions to differentiate between a non-aggressive versus aggressive police strategy.

(i) **No initial acts of aggression from police:** Due to the homogeneity of the crowd if there are no initial acts of aggression by the police, then there is no sensitivity to the parameters and no acts of aggression from either protesters or police occur. The tension dissipates with time and moderates eventually leave and the protest comes to an end. We classify this case as a **productive protest**.

(ii) **Single initial act of aggression from police:** A single act of aggression perpetrated by the police changes the dynamics significantly with this homogeneously moderate crowd. This could, for example, be a single arrest. Figure (3) illustrates a global sensitivity analysis to parameters $T_1, T_2, \theta, \tau_c$ and $v_c$. The top three and the first two bottom panels in Figure 3 illustrate the global sensitivity of solutions with initial data set to

$$v_1(0) = 0, v_2(0) = 1, u_1(0) = 0, u_2(0) = 500, \tau(0) = 2$$

and $p = 100$, while the protest lasts. The black line illustrates the average dynamics of each of the five unknowns. We observe, for example, that the AofA perpetrated by the police and the agitators increases on average. Moreover, the number of agitators sharply increases and then decays very quickly. Note that the dynamics here show little sensitivity to the parameters. The dynamics of the moderates will always decrease.

The bottom left graph illustrates a local sensitivity analysis. From the bottom right panel in Figure 3 we observe that the system is most sensitive to $T_1$ and $T_2$. However, the sensitivity is due to the effect that these parameters have on the number of acts of aggression. Clearly, the higher the $T_1$ and $T_2$ are the more acts that occur. We observe that the dynamics of the moderates and agitators are not very sensitive to the parameters. On the other hand, the number of AofA perpetrated by the police and by the agitators are very sensitive to $T_1$ and $T_2$. Here we have a situation where a minimally aggressive police presence can significantly increase the number of AofA.

| Parameter | Min value | Max value |
|---|---|---|
| $T_1$ | 0.001 | 0.2 |
| $T_2$ | 0.0001 | 0.01 |
| $\theta$ | 0.01 | 0.08 |
| $v_c$ | 0 | 10 |
| $\tau_c$ | 0 | 10 |

**Table 2:** Parameter ranges used for the global sensitivity analysis.

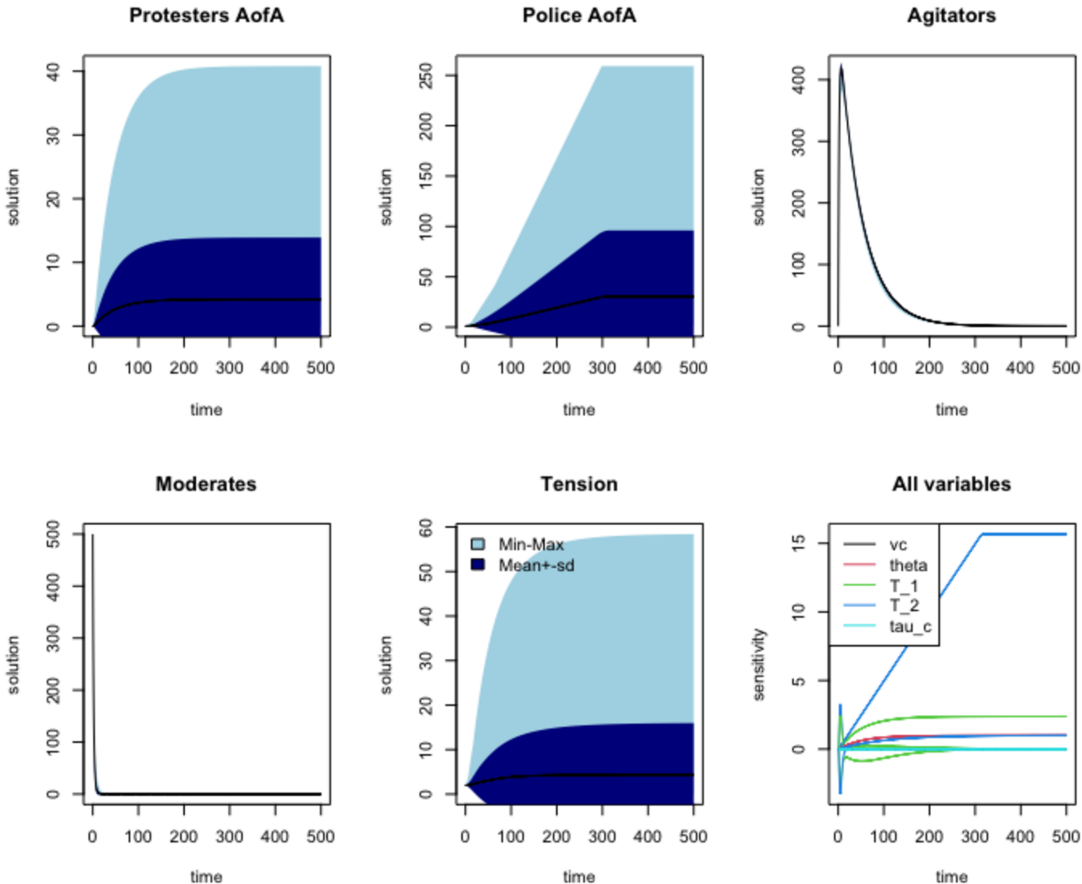

**Figure 3:** Local and global sensitivity analysis for case study 1(i) for parameters $T_1, T_2, \theta, \tau_c$ and $v_c$. The other parameters are set to $\epsilon = .02, \omega = .01$. The initial conditions are set to $v_1(0) = 0, v_2(0) = 1, u_1(0) = 0, u_2(0) = 500, \tau(0) = 2$ and $p = 100$ while the protest lasts.

#### 4.1.2 Case study 2: A heterogeneous crowd

We now look at what happens with an initial crowd that is heterogeneous. Specifically, we analyze what happens when twenty percent of the protesters are agitators at the start of the protest. We aim to analyze the evolution of the protest depending on when the police enter the picture.

(i) **Police presence from the start of protest:** We explore what happens when the police are present from the beginning. Figure 4 illustrates a global sensitivity analysis for this situation. The initial conditions for all runs are set to
$$v_1(0) = v_2(0) = 0, u_1(0) = 100, u_2(0) = 400, \tau(0) = 2$$
and $p(t) = 100$ for all times when there are more than one protesters. In this situation, we observe that the dynamics of the agitators and moderates are much more sensitive to the parameters. We see a maximum of 30 AofA from protesters and around 400 AofA from the police. Although the averages are around 2.5 and 10 AofA for protesters and police, respectively. Unlike the dynamics of the agitators in Case Study 1 (ii) where the agitators always increased, in this case study we observe that on average the number of agitators actually decreases. In the scenarios where they increase, they increase to a smaller number than that observed in Figure 3.

(ii) **Late entrance of the police:** Next, we consider what happens when the police show up later in the protest. We maintain the same initial conditions: $v_1(0) = v_2(0) = 0, u_1(0) = 100, u_2(0) = 400, \tau(0) = 2$. However, we switch the police presence function to $p = 0$ for $0 < t < 10$ and $p = 100$ for the remainder of the protest (until there is less than one protester). That is, there is no deterrence for agitators during the first ten units of time. The global sensitivity analysis for this case study is illustrated in Figure 5. Compared to the previous case study, we see an increased number of AofA both for police and protesters, from an average of 10 to around 50 for the police and from 2.5 to around 15 for protesters.

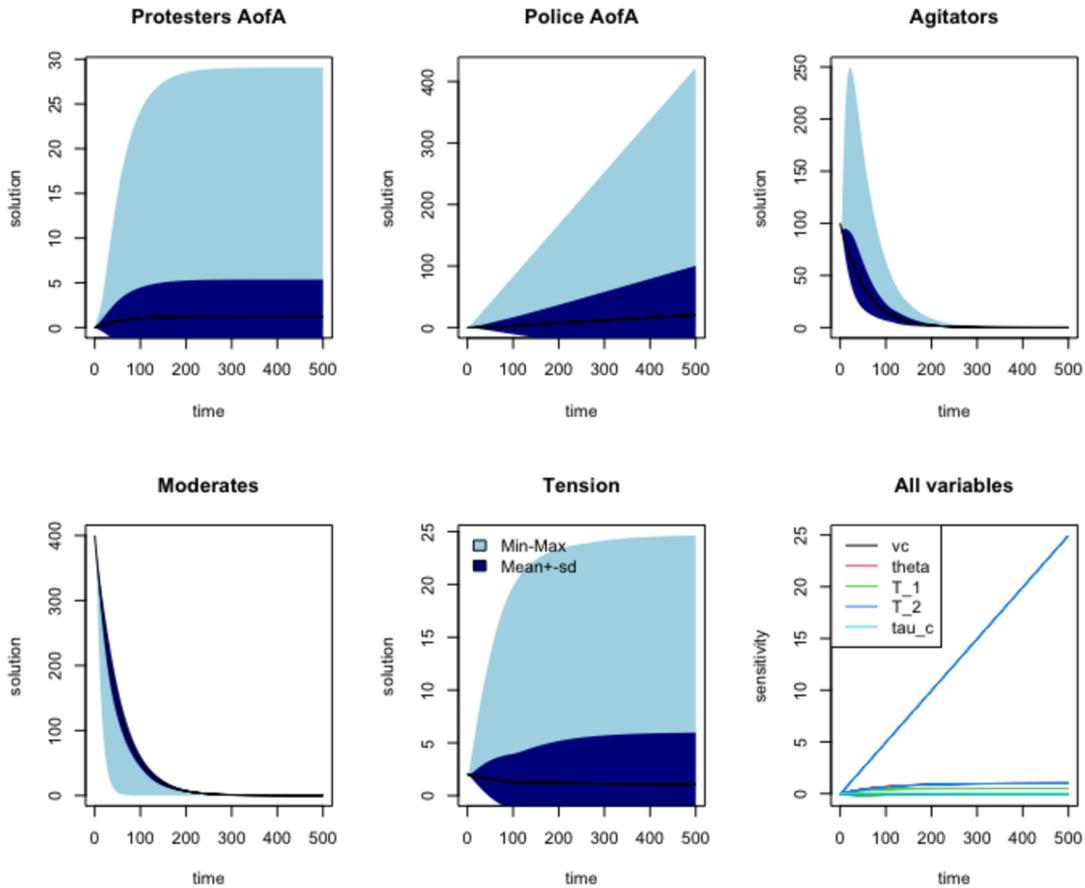

**Figure 4:** Local and global sensitivity analysis for case study 2(i) for parameters $T_1, T_2, \theta, \tau_c$ and $v_c$. The other parameters are set to $\epsilon = .02$ and $\omega = .01$. The initial conditions are set to $v_1(0) = 0, v_2(0) = 0, u_1(0) = 100, u_2(0) = 400, \tau(0) = 2$ and $p = 100$ for the duration of the protest.

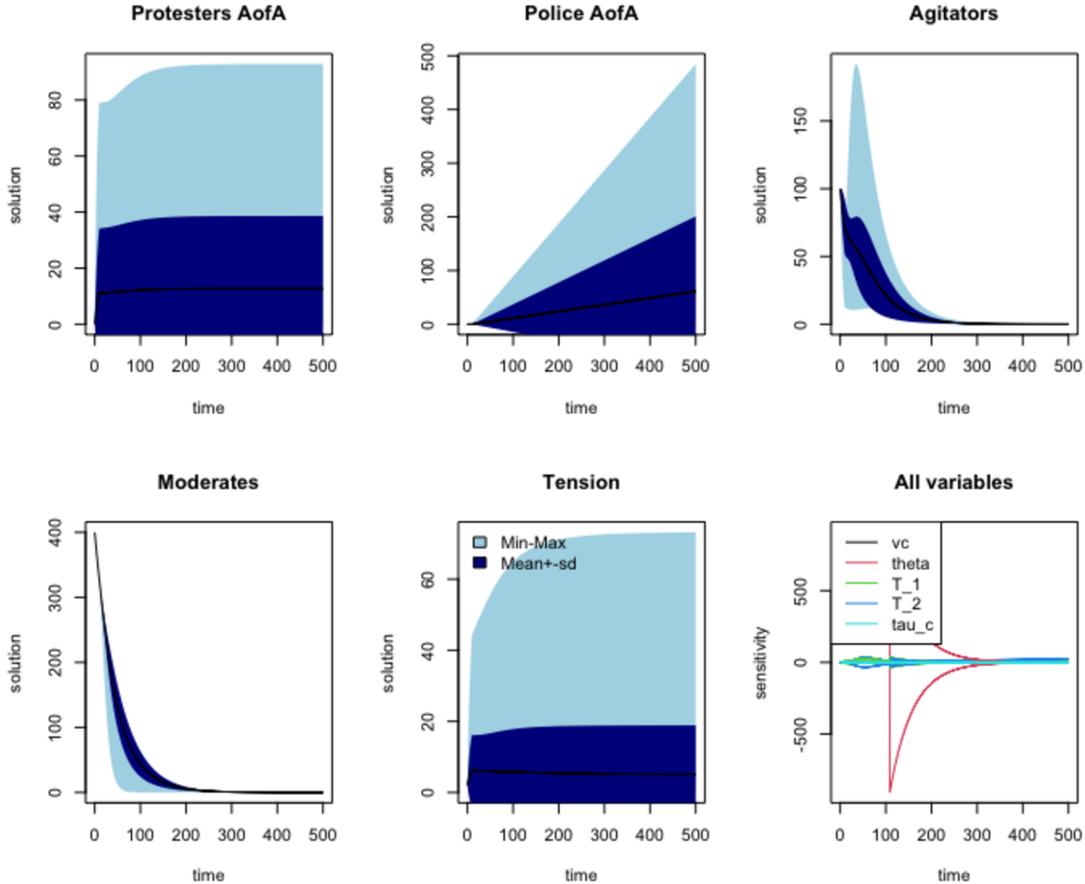

**Figure 5:** Local and global sensitivity analysis for case study 2(ii) for parameters $T_1, T_2, \theta, \tau_c$ and $v_c$. The other parameters are set to $\epsilon = .02$ and $\omega = .01$. The initial conditions are set to $v_1(0) = 0, v_2(0) = 0, u_1(0) = 100, u_2(0) = 400, \tau(0) = 2$ and $p = 100$ after $t = 10$ and for the remainder of the protest.

## 4.2 Dynamics of the protest

The two different police management models we are considering lead to different parameters. Note that in the escalated force model there is little to no tolerance for disruptions from the protesters and the police are quick to use force and arrests. This can be incorporated by using a low value for $v_c$. Moreover, it can also be incorporated by a larger value of $\theta$, however we will not explore the effect of this parameter in the present paper. On the other hand, in the negotiated management model, police only use force and arrests as a last resource, leading to higher values of $v_c$ (and lower values of the parameter $\theta$). In this section we fix $f_3(z) = \frac{1}{10}\mathbb{1}_{z>5}$.

### 4.2.1 Heterogeneous crowds

In this section, we perform a thorough exploration of the effect that $\tau_c$ and $v_c$ have on the dynamics of the system depending on the ratio of agitators and moderates. To achieve this we ran a large number of simulations where we fix all parameters with the exception of $\tau_c$ and $v_c$. In all of the simulations that are discussed in this section, we fix the values:

$$v_1(0) = 0, \ v_2(0) = 0, \ \tau(0) = 5, \text{and } N = 500$$

and vary the ratio of agitators to moderates.

Figure 6 contains two heat maps illustrating in panel (a) the total number of AofA perpetrated by the police and in panel (b) the total number of AofA perpetrated by the protesters. Figure 6a illustrates a clear phase transition between a large number of AofA and zero. The boundary at $\tau_c = 5$ is the result of the initial tension being set to $\tau(0) = 5$. Thus, when $\tau_c > 5$ there are no AofA committed by the protesters (as seen in Figure 6b). Since $v_1 \equiv 0$, this implies $v_2 \equiv 0$, except in the case when $v_c = 0$ in which case there is some AofA committed by the police. This is the thin aqua-blue line seen on the $v_c = 0$ axis. We see that the boundary between aggression and no aggression shifts left as $\tau_c$ decreases. Interestingly, this shows that when a crowd has a lower triggering point for committing AofA the police should be even more tolerant if they want to minimize how many AofA are committed. The shift from a high number of AofA to zero at the $\tau_c = 0$ axis occurs when $v_c = 9$ because the max number of AofA committed by protesters is nine. Shifting to Figure 6b we see that to the right of the phase transition boundary observed in Figure 6a there are still some AofA committed by the protesters. However, we see a shift to smaller numbers as $\tau_c$ increases.

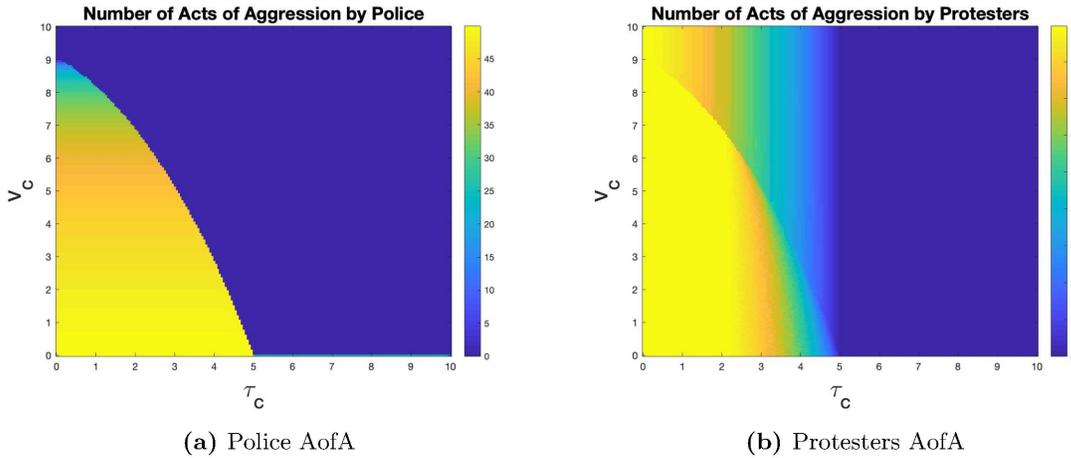

(a) Police AofA      (b) Protesters AofA

**Figure 6:** The number of acts of aggression perpetrated by the (a) police and (b) protesters as a function of $\tau_c$ and $v_c$. The ratio of agitators to the total number of protesters is taken to be $A/N = .2$. The parameters used here are $T_1 = .1, T_2 = .01, \theta = .2, \omega = .01, \epsilon = .01$. The total number of protesters is 500 and $p = 100$ while $u_1 + u_2 > 1$.

The heat maps illustrated in Figure 7 are the results of simulations with the exact same parameters and initial conditions as those used to generate Figure 6, except the initial ratio of agitators to the total number of protesters increased from $A/N = .2$ in Figure 6 to $A/N = .4$ in Figure 7. As expected, we see in increase in the number of AofA for both protesters and police. Moreover, for a given value of $\tau_c$ the regions with a positive number of AofA perpetrated by the police have increased. That is, the larger the number of agitators, the higher the police tolerance ought to be to avoid violence. An interesting feature that is observed in Figure 7b is that there is another phase transition between the total number of AofA perpetrated by the protesters (on the bottom left corner) as $v_c$ increases. There is a sharp drop in the number of AofA around $v_c = 1.1$.

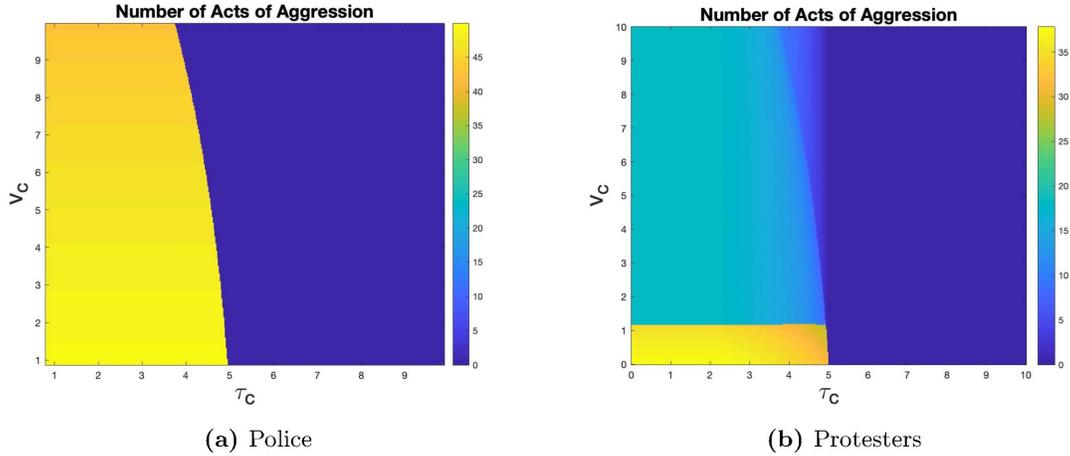

**(a)** Police        **(b)** Protesters

**Figure 7:** The number of acts of aggression perpetrated by the (a) police and (b) protesters as a function of $\tau_c$ and $v_c$. The ratio of agitators to the total number of protesters is taken to be $A/N = .4$. The same parameters are used as those given in Figure 6.

In Figure 8 we have increased the number of agitators to half of the total protesters. One can observe that the horizontal phase transition has shifted to around $v_c = 4.8$.

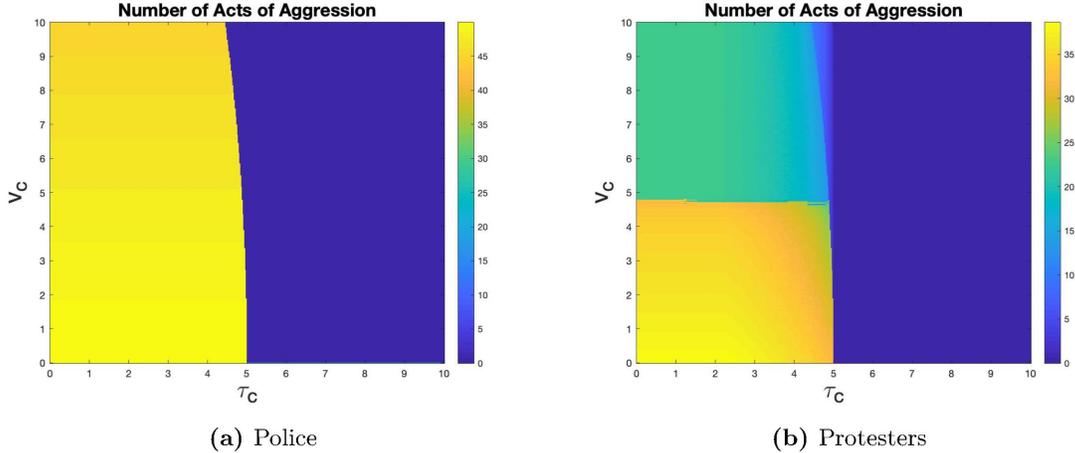

**(a)** Police        **(b)** Protesters

**Figure 8:** The number of acts of aggression perpetrated by the (a) police and (b) protesters as a function of $\tau_c$ and $v_c$ when the ratio of agitators to the total number of protesters is $A/N = .5$. The parameters used here are the same used for the simulations in Figure 6.

Some key takeaways from these simulations are that there is a clear phase transition between the number of AofA committed by the police, going from zero to a significant number. When $\tau_c$ is higher, $v_c$ needs to be higher in order for the AofA perpetrated by the police to remain. As the number of initial agitators increases we also see a second phase transition appear, where there is a small, but clear drop between the number of AofA committed by the protesters.

### 4.3   Entrance time versus heterogeneity of the crowd

In this section, we analyze the effect that police entrance time has on the dynamics of the protests. In particular, we are interested in how the total number of AofA perpetrated both by the police and protesters changes when we shift the police entrance time and the composition of the crowd. All of the simulations used to generate the heat maps in this section used the following initial conditions: $v_1(0) = v_2(0) = 0$. Moreover, $N = 500$ and $p = 100$. Here we vary the number of initial agitators and the police entrance time.

Figure 9 illustrates the number of AofA by the police and protesters when $\tau_c = 5$, $v_c = 15$, $T_1 = .1$ and $T_2 = .001$. The first thing to note is the sharp phase transition seen in Figure 9a. As the number of initial agitators increases, the entrance time of the police matters more. That is, as the initial number of agitators increases the sooner the police must enter to reduce the number of AofA perpetrated by the police. However, after around $u_3(0) = 350$, the entrance time does not matter as much for the total police AofA. We see a similar pattern in a number of AofA perpetrated by the protesters, see Figure 9b. However, there is a gradual change in the number of AofA perpetrated by the protesters. In fact, in the top right corner, we see a concentration of high numbers of AofA. This implies that police showing up earlier reduces the total trauma.

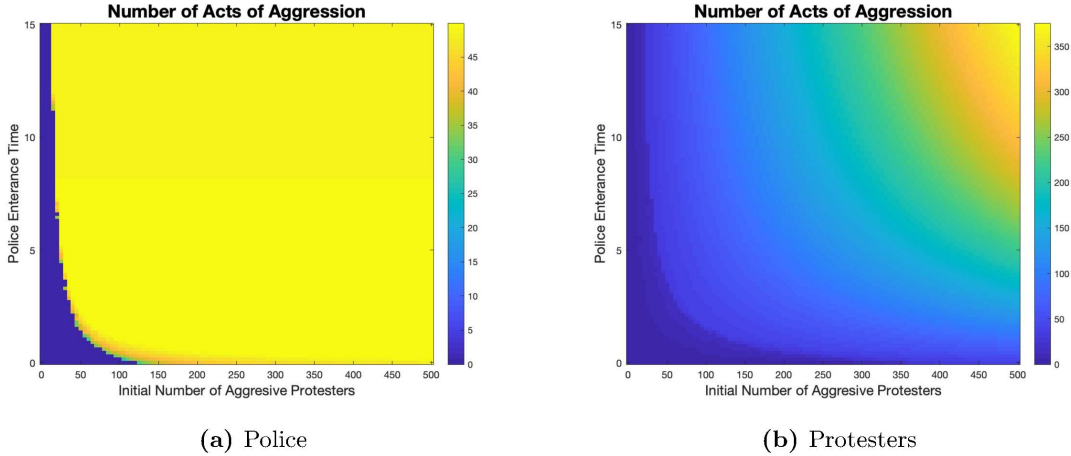

(a) Police

(b) Protesters

**Figure 9:** Heat maps for the total number of AofA perpetrated by the police in (a) and protesters in (b). The parameters used are $\tau_c = 5$, $v_c = 15$, $T_1 = .1$ and $T_2 = .001$.

Note that we have taken $T_1 = .1$ for the simulations that produced the heat maps illustrated in Figure 9. We can think of $T_1$ as a parameter that modulates the effects of deterrence of AofA by the police presence versus that of procedural justice. We can compare these results to those seen in Figure 10 where all things are equal with the exception that $T_1 = .01$. In this case, the deterrence effect is much stronger than in the one seen in Figure 9. The patterns observed are the same; however, we do note a shift to the upper right. Specifically, in Figure 10a we see that there is more wiggle room for mitigating police AofA in terms of the entrance time of the police. When $T = .1$ and the initial number of agitators was 500, the entrance time did not matter as the same number of AofA occur independently of when the police arrived. On the other hand, when $T = .01$ the police entrance time can be a little bit delayed without increasing the number of AofA by the police.

Finally, Figure 11 illustrates the case when $T_1 = .5$, which leads to a case where the procedural justice effect is much stronger than the deterrence effect of police presence. As expected, this leads to a larger region of parameters that lead to a significant number of AofA committed by the police, see Figure 11a. Moreover, we see that the region with the maximum number of AofA committed by the protesters (top right corner - in bright yellow) is stretched downwards significantly. Naturally, the number of AofA committed by protesters is much larger.

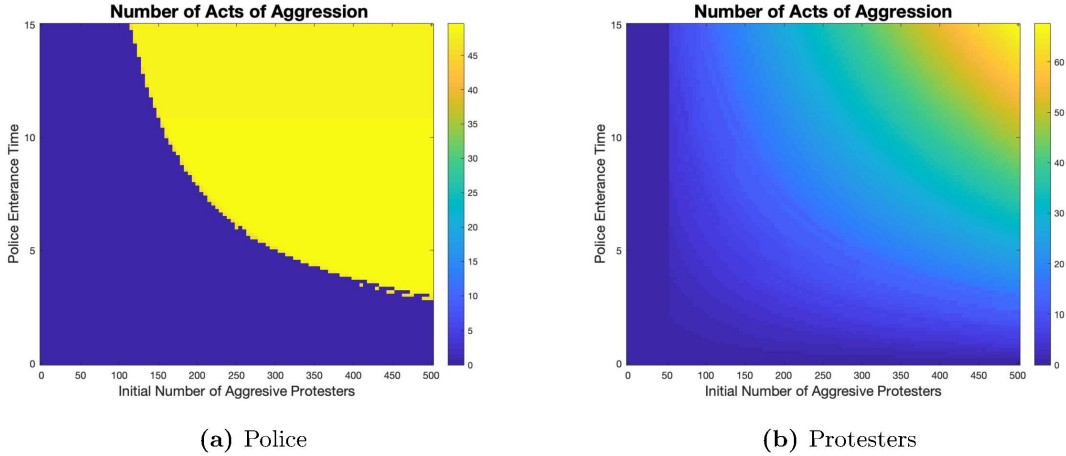

**(a)** Police        **(b)** Protesters

**Figure 10:** The number of acts of aggression perpetrated by the (a) police and (b) protesters as a function of the total number of initial agitators and police entrance times. The parameters used for all simulations are $\tau_c = 5$, $v_c = 15$, $T_1 = .01$ and $T_2 = .001$.

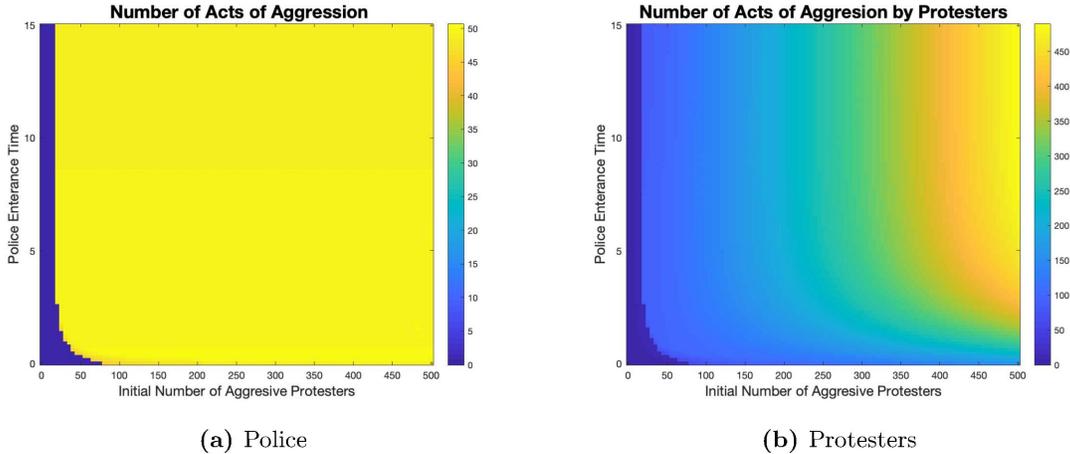

**(a)** Police        **(b)** Protesters

**Figure 11:** The number of acts of aggression perpetrated by the (a) police and (b) protesters as a function of the total number of initial agitators and police entrance times. The parameters used for all simulations are $\tau_c = 5$, $v_c = 15$, $T_1 = .5$ and $T_2 = .001$.

## 5 Discussion and conclusions

In this work, we presented a two-pronged approach to understanding the effect that different police protest management models have on how protests evolve. In particular, we aimed to understand the effect that different management models have on trauma, which can occur from things such as violence perpetrated by protesters and police, as well as arrests. Our statistical approach made use of event data provided by ACLED. From a time series analysis, we learned that kinetic impact projectiles (KIPs) usage is associated with increased numbers of protests in the subsequent days. Moreover, the use of KIPs is associated with an increased number of deaths in subsequent days. In fact, the use of KIPs was a better predictor for deaths than the number of protests. There is widespread agreement that police should use the minimum amount of force necessary to achieve their goals and that anything beyond that is excessive force. Our work adds to the viewpoint that the use of KIPs is definitely not the minimum. Even if "less-lethal" tactics are needed, KIPs should not be used because they are more harmful than the other "less-lethal" options.

The key takeaway from our dynamical modeling approach is that understanding the heterogeneity of the crowd is essential. An aggressive strategy, such as the use of the escalated management model, or a simple accidental act of aggression perpetrated by the police, can lead to escalation of trauma. In particular, a peaceful crowd would lead to no trauma (and reduced social tension), but an improper police response could lead to significant trauma. We conclude that the negotiated management model is always a better model, independent of the type of crowd. In fact, the lower the threshold for acts of aggression from the crowd, the more tolerant the police should be of aggressive actions by the crowd. This could significantly lead to less group trauma. Another interesting takeaway is that if a crowd has some agitators, police presence can help reduce the total trauma, provided the right policing strategy is used. Early police presence in combination with a low threshold for violence can lead to even more trauma.

The benefit of the dynamical approach used here is that we can provide a mathematical framework for protest dynamics based on modern theories of compliance and deterrence theory and crowd psychology. This framework is especially useful given the lack of deep data necessary to fully understand the influence of police on protester behavior and trauma. This mathematical framework can help us quantify the group trauma that is experienced in protests depending on police behavior. The results here are semi-quantitative, in the sense that we cannot determine what parameter regime certain events are in. We urge for the collection of detailed data and data availability, which needs to be made by the police, to be able to obtain quantitative results that can be verified.

# Acknowledgments

We are grateful to the Institute for Computational and Experimental Research in Mathematics (ICERM) where this project was begun as part of a program on Data Science and Social Justice. We also want to thank the ACLED team for making these data publicly available. Rodríguez was partially funded by NSF-DMS 2042413 and AFOSR MURI grant number FA9550-22-1-0380. We would like to thank Drs. Nicole Koulisis and Stavros Moysidis for motivating this work.